\documentclass[aps,prc,twocolumn,fleqn]{revtex4}

\usepackage{amsmath,amssymb,amsopn}
\usepackage{epsfig}
\usepackage{xcolor}

\begin{document}

\title{Transverse Energy Density Fluctuations in the Color Glass Condensate Model}

\author{Berndt M\"uller}
\affiliation{Department of Physics, Duke University, Durham, NC 27708}
\author{Andreas Sch\"afer}
\affiliation{Institut f\"ur Theoretische Physik, Universit\"at Regensburg, 
93040 Regensburg, Germany}

\date{\today}

\begin{abstract}
We calculate the transverse correlation of fluctuations of the deposited energy density in nuclear collisions in the framework of the Gaussian color glass condensate model.
\end{abstract}

\maketitle

\section{Event-by-event fluctuations}

The event-by-event fluctuations of the transverse emission pattern of hadrons in high-energy collisions of identical heavy nuclei have recently attracted much interest experimentally \cite{Alver:2007qw,Sorensen:2008zk,ALICE:2011vk,Adare:2011tg,ATLAS:2011hf,Sorensen:2011fb,Appelt:2011mw} and theoretically \cite{Miller:2003kd,Broniowski:2007ft,Andrade:2008xh,Shuryak:2009cy,Alver:2010gr,Petersen:2010cw,Alver:2010dn,Staig:2011as,Holopainen:2010gz,Qin:2010pf,Schenke:2010rr,Qiu:2011iv,Bhalerao:2011yg,Qin:2011uw,deSouza:2011rp}. When averaged over collision events, the azimuthal angular distribution of emitted hadrons around the beam axis is symmetric with respect to the plane perpendicular to the impact parameter vector ${\bf b}$ between the two nuclei [Au+Au or Cu+Cu at the Relativistic Heavy Ion Collider (RHIC) or Pb+Pb at the Large Hadron Collider (LHC)]. The event averaged angular distribution
\begin{equation}
\frac{dN}{d^2p_T} = \frac{dN}{\pi dp_T^2} \Big(1 + \sum_{n=1}^\infty v_n(p_T) \cos(\phi_p) \Big) ,
\end{equation}
where $\phi_p$ is the angle between ${\bf p}_T$ and ${\bf b}$, is therefore completely characterized by the even Fourier coefficients $v_n$. The dominant coefficient, $v_2$, is called elliptic flow. 

Owing to quantum fluctuations in the density distributions of the colliding nuclei and finite particle number effects on the distribution of emitted particles, the left-right symmetry is broken in individual collision events. The angular distribution can then be written in the form
\begin{equation}
\frac{dN}{d^2p_T} = \frac{dN}{\pi dp_T^2} \Big(1 + \sum_{n=1}^\infty v_n(p_T) \cos(\phi_p+\psi_n) \Big) ,
\end{equation}
where $\psi_n$ describes the tilt angle of the ``event plane'' for each Fourier coefficient with respect to the reaction plane defined by the vector ${\bf b}$. For even $n$, $\psi_n$ is peaked around zero; for odd $n$, $\psi_n$ is randomly distributed.  The dominant odd coefficient, $v_3$, is known as triangular flow. The event averages of the amplitude of the coefficients $v_n$ are found to be constant over a rather large pseudorapidity range ($|\eta| \leq 2$) in Pb+Pb collisions at the LHC \cite{ATLAS:2011hf}, indicating an approximately boost invariant origin.

The main dynamical source of event-by-event fluctuations in the coefficients $v_n$ are believed to be nearly boost invariant fluctuations in the transverse distribution of the energy density at the beginning of the hydrodynamic expansion of the quark-gluon plasma formed in the nuclear collisions. The geometric anisotropy of these fluctuations then translates in an anisotropic transverse collective flow pattern, which manifests itself in anisotropic particle emission. In the color glass condensate model of energy deposition there are two obvious sources of fluctuations in the deposited energy density.  One is geometric fluctuations of the position of nucleons in the colliding nuclei at the moment of impact, leading to transverse fluctuations in the density of field generating color charges. This mechanism has been studied widely and is usually described geometrically by the Monte-Carlo Glauber model \cite{Drescher:2006ca}. The transverse correlation length of the fluctuations generated by this mechanism will be of the order of the nucleon radius.

The other source of energy density fluctuations are fluctuations in the color field strength for a given density of color charges. This mechanism has not been investigated quantitatively up to now. The transverse correlation length generated by color field fluctuations will be dictated by the single scale governing the physics of the color glass condensate, the saturation scale $Q_s$. Since $Q_s^{-1}$ is much smaller than the nucleon radius, the color field fluctuations can be expected to govern the microscopic structure of the transverse energy density fluctuations, which then is modulated on longer transverse scales by fluctuations in the nucleon density in the colliding nuclei.

Here we calculate the transverse correlation function of the deposited energy density in nuclear collisions in the framework of the Gaussian approximation to the color glass condensate originally proposed by McLerran and Venugopalan \cite{McLerran:1993ni,McLerran:1993ka}. In Section II we derive the two-point correlator of the energy density following the collision of two color glass condensates. In Section III we evaluate the resulting integrals and present numerical results for the correlation function for a representative choice of parameters.

\section{Energy density fluctuations}

In the Gaussian random source approximation to the color glass condensate model of small--$x$ gluon structure of atomic nuclei \cite{McLerran:1993ni,McLerran:1993ka} the probability distribution of color charge density $\rho^a({\bf x})$ in the transverse plane is assumed to be of the form
\begin{equation}
\label{eq:Gauss-rho}
P[\rho^a] = \exp\left( - \frac{1}{g^2\mu^2} \int d^2x\, \rho^a({\bf x})^2 \right) .
\end{equation}
Here $\mu^2$ represents the area density of color charges in the colliding nuclei, and $Q_s = g\mu$ is called the saturation scale, because it represents the scale at which the small--$x$ evolution of the gluon density becomes nonlinear due to saturation effects \cite{JalilianMarian:1997gr,Mueller:1999wm}. Owing to the independent contributions of several nucleons to the color field, the Gaussian approximation is expected to provide a good description to the color source distribution in colliding nuclei at small $x$ \cite{Kovchegov:1996ty}. In the light-cone gauge, the Gaussian color charge distribution translates into a Gaussian distribution of transverse gauge field strengths. Here we will follow the work of Lappi  \cite{Lappi:06}. 

To calculate the initial state density fluctuations 
\begin{equation}
\left\langle \varepsilon({\bf x}) \varepsilon({\bf y}) \right\rangle 
-\left\langle \varepsilon({\bf x}) \right\rangle \left\langle\varepsilon({\bf y}) \right\rangle ,
\end{equation}
where ${\bf x,y}$ denote vectors in the transverse plane, we start from the expression for the deposited energy density of the gauge field given in Eq.(10) in ref.~\cite{Lappi:06}:
\begin{equation}
\varepsilon({\bf x}) = \frac{1}{4} F^c_{ij}({\bf x})F^c_{ij}({\bf x}) +2 A^{\eta c}({\bf x}) A^{\eta c}({\bf x})  
\end{equation}
with transverse vector indices $i,j,m,n,...=1,2$. Following the collision, the field strength tensor in the region between the receding nuclei only receives contributions from the mixed terms, as the color field of each individual nucleus is a pure gauge and the field strength tensor of each individual nucleus is thus zero outside the nuclear volume:
\begin{eqnarray}
F^c_{ij}({\bf x}) &=& gf_{abc} \Big( A_i^a(1;{\bf x}) A^b_j(2;{\bf x}) 
\nonumber \\
&& +  A_i^a(2;{\bf x}) A^b_j(1;{\bf x}) \Big) 
\\
A^{\eta c}({\bf x}) A^{\eta c}({\bf x}) &=& 
\frac{g^2}{4}f_{abc}f_{a'b'c}
\\
&\times&A^a_i(1;{\bf x})A^b_i(2;{\bf x})A^{a'}_j(1;{\bf x})A^{b'}_j(2;{\bf x}) .
\nonumber
\end{eqnarray}  
Here ``1'' and ``2'' denote the gauge fields carried by nucleus 1 and 2, respectively. The field correlator in the color glass condensate model is given by
\begin{widetext} 
\begin{eqnarray}
\Big\langle A^a_i(n;{\bf x}) A^b_j(m;{\bf y}) \Big\rangle
&=&
\frac{1}{2} \Big\langle A^a_i(n;{\bf x}) A^b_j(m;{\bf y}) \Big\rangle 
~+~
\frac{1}{2}\Big\langle A^b_j(m;{\bf y}) A^a_i(n;{\bf x}) \Big\rangle 
\nonumber \\
&=& 
\delta_{mn}\delta_{ab}~\int \frac{d^2p}{(2\pi)^2} ~ \cos[{\bf p}\cdot ({\bf x}-{\bf y})]
~\frac{p_ip_j}{{\bf p}^2} ~G(|{\bf p}|) 
\end{eqnarray}
where $G(|{\bf p}|)$ is the Fourier transform of the function
\begin{equation}
G(|{\bf x}|) = \frac{4}{g^2 N |{\bf x}|^2} 
\left[ 1-\exp\left( \frac{g^2N}{8\pi} g^2\mu^2 |{\bf x}|^2\ln(\Lambda|{\bf x}|) \right) \right]
~\Theta(1-\Lambda|{\bf x}|)
\label{eq:G(x)}
\end{equation}
with the IR cut-off parameter $\Lambda$. It is convenient to decompose the momentum quadrupole tensor as follows:
\begin{equation}
p_ip_j = \frac{p_1^2+p_2^2}{2}\,\delta_{ij} +  \frac{p_1^2-p_2^2}{2}\,\sigma^3_{ij}
+ p_1p_2\,\sigma^1_{ij} ,
\end{equation}
where $\sigma^1,\sigma^3$ are the familiar Pauli matrices. We thus obtain
\begin{eqnarray}
\Big\langle A^a_i(n;{\bf x}) A^b_j(m;{\bf y}) \Big\rangle  
&=& \frac{1}{2}\, \delta_{mn}\delta_{ab}~\int \frac{d^2p}{(2\pi)^2}
\Big( \cos[p_1 (x_1-y_1)]\cos[p_2 (x_2-y_2)] ~\delta_{ij} ~G(|{\bf p}|)
\nonumber \\
&& 
+ \cos[p_1 (x_1-y_1)]\cos[p_2 (x_2-y_2)] ~\sigma^3_{ij}~\frac{p_1^2-p_2^2}{{\bf p}^2}~G(|{\bf p}|)
\nonumber \\
&& 
- \sin[p_1 (x_1-y_1)]\sin[p_2 (x_2-y_2)] ~ \sigma^1_{ij} ~\frac{p_1p_2}{{\bf p}^2}~G(|{\bf p}|) \Big)
\nonumber\\
&=&
\frac{1}{2} \, \delta_{mn}\delta_{ab}~ \Big(\delta_{ij} D({\bf x}-{\bf y}) 
+ \sigma^3_{ij} E({\bf x}-{\bf y}) - \sigma^1_{ij} F({\bf x}-{\bf y})\Big) 
\equiv \delta_{mn}\delta_{ab}~ S_{ij}({\bf x}-{\bf y}) .
\end{eqnarray}
For later use, we will note the values of the individual correlation functions $D, E, F$ at the origin:
\begin{eqnarray}
D(0) &=& \int \frac{d^2p}{(2\pi)^2} ~G(|{\bf p}|) = \lim_{|{\bf x}|\to 0} G(|{\bf x}|) ;
\qquad
E(0) = F(0) = 0 .
\end{eqnarray}
The expression for $D(0)$ diverges logarithmically for the function $G(|{\bf x}|)$ given in (\ref{eq:G(x)}), if the gauge coupling $g$ is taken as a constant. However, as pointed out by Kovchegov and Weigert \cite{Kovchegov:2007vf}, the infrared divergence can be removed by including effects from the running of the coupling constant by means of the substitution
\begin{equation}
g^4 \longrightarrow g^2(\mu^2) g^2(1/|{\bf x}|^2) 
\end{equation}
in the exponent of (\ref{eq:G(x)}). The specific structure of this substitution, sometimes called the ``triumvirate'' structure of the running coupling, is motivated by the form of next-to-leading order corrections to the small--$x$ evolution of the BFKL kernel in the color dipole approach to parton saturation \cite{Kovchegov:2006wf}.

We first evaluate the expectation value of the deposited energy density:
\begin{eqnarray}
\langle \varepsilon({\bf x})\rangle &=& \frac{g^2}{2}f_{abc}f_{a'b'c} 
\Big\langle A_i^a(1;{\bf x}) A^b_j(2;{\bf x}) A_i^{a'}(1;{\bf x}) A^{b'}_j(2;{\bf x})
+  A_i^a(1;{\bf x}) A^b_j(2;{\bf x}) A_i^{a'}(2;{\bf x}) A^{b'}_j(1;{\bf x}) \Big\rangle
\nonumber \\
&& + \frac{g^2}{2}f_{abc}f_{a'b'c}
\Big\langle A^a_i(1;{\bf x})A^b_i(2;{\bf x})A^{a'}_j(1;{\bf x})A^{b'}_j(2;{\bf x})\Big\rangle
\nonumber \\
&=&
\frac{g^2}{2}f_{abc}f_{a'b'c}D(0)^2 
\Big( \delta_{aa'}\delta_{bb'} +
\delta_{ab'}\delta_{ba'}/2 
+\delta_{aa'}\delta_{bb'}/2 \Big)
\nonumber \\
&=& 
\frac{g^2}{2} N(N^2-1)D^2(0) ,
\end{eqnarray}
recovering the result given in Eq.~(14) of ref.~\cite{Lappi:06}.
 
Next we evaluate the two-point correlator of the energy density:
\begin{eqnarray}
\langle \varepsilon({\bf x})\varepsilon({\bf y})\rangle 
&=& \frac{g^4}{4}f_{abc}f_{a'b'c} f_{efd}f_{e'f'd} \,
\Big\langle \Big( A_i^a(1;{\bf x}) A^b_j(2;{\bf x}) A_i^{a'}(1;{\bf x}) A^{b'}_j(2;{\bf x})
\nonumber \\
&&+ A_i^a(1;{\bf x}) A^b_j(2;{\bf x}) A_i^{a'}(2;{\bf x}) A^{b'}_j(1;{\bf x}) 
+ A^a_i(1;{\bf x})A^b_i(2;{\bf x})A^{a'}_j(1;{\bf x})A^{b'}_j(2;{\bf x}) \Big)
\nonumber \\
&&\times \Big( A_m^e(1;y) A^f_n(2;y) A_m^{e'}(1;y) A^{f'}_n(2;y)
+ A_m^e(1;y) A^f_n(2;y) A_m^{e'}(2;y) A^{f'}_n(1;y) 
\nonumber\\
&& + A^e_m(1;y)A^f_m(2;y)A^{e'}_n(1;y)A^{f'}_n(2;y) \Big)
\Big\rangle
\end{eqnarray}
We again make use of the fact that only correlators among fields in the same nucleus are non-zero, which 
allows us to suppress the labels 1 and 2:
\begin{eqnarray}
\langle \varepsilon({\bf x})\varepsilon({\bf y})\rangle 
&=& \frac{g^4}{4}f_{abc}f_{a'b'c} f_{efd}f_{e'f'd} \, 
\Big( \Big\langle A_i^a({\bf x}) A_i^{a'}({\bf x})A_m^e({\bf y}) A_m^{e'}({\bf y})\Big\rangle
 \Big\langle A_j^b({\bf x}) A_j^{b'}({\bf x})A_n^f({\bf y}) A_n^{f'}({\bf y})\Big\rangle
\nonumber \\
&&+ \Big\langle A_i^a({\bf x}) A_i^{a'}({\bf x})A_m^e({\bf y}) A_n^{f'}({\bf y})\Big\rangle
\Big\langle A_j^b({\bf x}) A_j^{b'}({\bf x})A_n^f({\bf y}) A_m^{e'}({\bf y})\Big\rangle
\nonumber \\
&&+ \Big\langle A_i^a({\bf x}) A_i^{a'}({\bf x})A_m^e({\bf y}) A_n^{e'}({\bf y})\Big\rangle
\Big\langle A_j^b({\bf x}) A_j^{b'}({\bf x})A_m^f({\bf y}) A_n^{f'}({\bf y})\Big\rangle
\nonumber \\
&&+ \Big\langle A_i^a({\bf x}) A_j^{b'}({\bf x})A_m^e({\bf y}) A_m^{e'}({\bf y})\Big\rangle
\Big\langle A_j^b({\bf x}) A_i^{a'}({\bf x})A_n^f({\bf y}) A_n^{f'}({\bf y})\Big\rangle
\nonumber \\
&&+ \Big\langle A_i^a({\bf x}) A_j^{b'}({\bf x})A_m^e({\bf y}) A_n^{f'}({\bf y})\Big\rangle
\Big\langle A_j^b({\bf x}) A_i^{a'}({\bf x})A_n^f({\bf y}) A_m^{e'}({\bf y})\Big\rangle
\nonumber \\
&&+ \Big\langle A_i^a({\bf x}) A_j^{b'}({\bf x})A_m^e({\bf y}) A_n^{e'}({\bf y})\Big\rangle
\Big\langle A_j^b({\bf x}) A_i^{a'}({\bf x})A_m^f({\bf y}) A_n^{f'}({\bf y})\Big\rangle
\nonumber \\
&&+ \Big\langle A_i^a({\bf x}) A_j^{a'}({\bf x})A_m^e({\bf y}) A_m^{e'}({\bf y})\Big\rangle
\Big\langle A_i^b({\bf x}) A_j^{b'}({\bf x})A_n^f({\bf y}) A_n^{f'}({\bf y})\Big\rangle
\nonumber \\
&&+ \Big\langle A_i^a({\bf x}) A_j^{a'}({\bf x})A_m^e({\bf y}) A_n^{f'}({\bf y})\Big\rangle
\Big\langle A_i^b({\bf x}) A_j^{b'}({\bf x})A_n^f({\bf y}) A_m^{e'}({\bf y})\Big\rangle
\nonumber \\
&&+ \Big\langle A_i^a({\bf x}) A_j^{a'}({\bf x})A_m^e({\bf y}) A_n^{e'}({\bf y})\Big\rangle
\Big\langle A_i^b({\bf x}) A_j^{b'}({\bf x})A_m^f({\bf y}) A_n^{f'}({\bf y})\Big\rangle \Big)
\nonumber \\
&\equiv & \frac{g^4}{4}f_{abc}f_{a'b'c} f_{efd}f_{e'f'd} \, \sum_{\alpha=1}^9 M_\alpha .
\label{eq:epseps}
\end{eqnarray}
In the spirit of the Gaussian approximation we now factorize the correlators of four gauge fields into products of correlators among two gauge fields \footnote{We note that the Gaussian approximation for the transverse gauge field components is not an immediate consequence of the Gaussian distribution for color charges, Eq.~(\ref{eq:Gauss-rho}), because the connection between $\rho({\bf x})$ and $A_i^a({\bf x})$ is nonlinear \cite{McLerran:1993ka,Marquet:2010cf}. For a Gaussian source charge distribution, these nonlinearities will thus lead to corrections to the energy density correlator obtained here.}, e.~g.:
\begin{eqnarray}
\Big\langle A_j^b({\bf x}) A_i^{a'}({\bf x})A_n^f({\bf y}) A_m^{e'}({\bf y})\Big\rangle
&=&\Big\langle A_j^b({\bf x}) A_i^{a'}({\bf x})\Big\rangle \Big\langle A_n^f({\bf y}) A_m^{e'}({\bf y})\Big\rangle
+ \Big\langle A_j^b({\bf x}) A_n^{f}({\bf y})\Big\rangle \Big\langle A_i^{a'}({\bf x}) A_m^{e'}({\bf y})\Big\rangle
\nonumber \\
&& + \Big\langle A_j^b({\bf x}) A_m^{e'}({\bf y})\Big\rangle \Big\langle A_n^f({\bf y}) A_i^{a'}({\bf x})\Big\rangle
\end{eqnarray}

To proceed further we use the symmetry with respect to the color indices $e'$ and $f'$ to combine, e.~g.,
the second and third term in the large brackets of (\ref{eq:epseps}): 
\begin{eqnarray}
M_2 + M_3 &=& \Big\langle A_i^a({\bf x}) A_i^{a'}({\bf x})A_m^e({\bf y}) A_n^{f'}({\bf y})\Big\rangle
 \Big\langle A_j^b({\bf x}) A_j^{b'}({\bf x})\Big(A_n^f({\bf y}) A_m^{e'}({\bf y})-A_m^f({\bf y}) A_n^{e'}({\bf y})\Big)\Big\rangle
\end{eqnarray}
The second factor is easily shown to vanish:
\begin{eqnarray}
\Big\langle \cdots \Big\rangle &=&
\Big\langle A_j^b({\bf x}) A_n^f({\bf y}) \Big\rangle  
\Big\langle A_j^{b'}({\bf x}) A_m^{e'}({\bf y}) \Big\rangle
\Big\langle A_j^{b'}({\bf x}) A_n^f({\bf y}) \Big\rangle  
\Big\langle A_j^{b}({\bf x}) A_m^{e'}({\bf y}) \Big\rangle -(m\leftrightarrow n)  
\nonumber \\
&=&
\delta_{bf}\delta_{b'e'}\, S_{jn}({\bf x}-{\bf y}) S_{jm}({\bf x}-{\bf y}) 
+ \delta_{b'f}\delta_{be'}\, S_{jn}({\bf x}-{\bf y}) S_{jm}({\bf x}-{\bf y})
 -(m\leftrightarrow n)
= 0
\end{eqnarray}  

The same holds true for the forth and seventh term, $M_4$ and $M_7$.
After considerable algebra, the fifth, sixth, eighth and ninth terms combine to 
\begin{eqnarray}
M_5 + M_6 + M_8 + M_9
&=& \frac{g^4}{4}f_{abc}f_{a'b'c} f_{efd}f_{e'f'd} \,
 \Big\langle A_i^a({\bf x}) A_j^{a'}({\bf x})A_m^e({\bf y}) A_n^{f'}({\bf y})\Big\rangle
\nonumber \\
&&\times \Big\langle \left[A_i^b({\bf x}) A_j^{b'}({\bf x})-A_j^b({\bf x}) A_i^{b'}({\bf x}) \right]
 \left[A_n^f({\bf y}) A_m^{e'}({\bf y})-A_m^f({\bf y}) A_n^{e'}({\bf y})\right]  \Big\rangle
\nonumber \\
&=&\frac{g^4}{4}f_{abc}f_{a'b'c} f_{efd}f_{e'f'd} \,
 \Big(\delta_{ae}\delta_{a'f'}S_{im}S_{jn} + \delta_{af'}\delta_{a'e}S_{in}S_{jm}\Big)
\nonumber \\
&&\times \Big(\delta_{bf}\delta_{b'e'}2[S_{in}S_{jm}-S_{im}S_{jn}] 
+ \delta_{be'}\delta_{b'f}2[S_{im}S_{jn}-S_{in}S_{jm}]\Big)
\nonumber \\
&=&\frac{g^4}{16}N^2(N^2-1)
\left[ D({\bf x}-{\bf y})^2 + E({\bf x}-{\bf y})^2 + F({\bf x}-{\bf y})^2 \right]^2 ,
\end{eqnarray}
\end{widetext}
where we made use of the relation 
\begin{equation}
f_{abc}f_{a'b'c}f_{a'bd}=\frac{N}{2}f_{ab'd}
\end{equation}
which follows from the Jacobi identity.

Finally we evaluate the first term:
\begin{widetext}
\begin{eqnarray}
M_1 &=& \frac{g^4}{4}f_{abc}f_{a'b'c} f_{efd}f_{e'f'd} \,
\Big\langle A_i^a({\bf x}) A_i^{a'}({\bf x})A_m^e({\bf y}) A_m^{e'}({\bf y})\Big\rangle
\Big\langle A_j^b({\bf x}) A_j^{b'}({\bf x})A_n^f({\bf y}) A_n^{f'}({\bf y})\Big\rangle
\nonumber \\
&=& \frac{g^4}{4}N^2(N^2-1)^2 D(0)^4
\nonumber \\
&+& \frac{g^4}{2}N^2(N^2-1) D(0)^2
\left[ D({\bf x}-{\bf y})^2 + E({\bf x}-{\bf y})^2 + F({\bf x}-{\bf y})^2 \right]
\nonumber \\
&+& \frac{3g^4}{8}N^2(N^2-1) 
\left[ D({\bf x}-{\bf y})^2 + E({\bf x}-{\bf y})^2 + F({\bf x}-{\bf y})^2 \right]^2 .
\end{eqnarray}
Combining these equations we finally obtain
\begin{eqnarray}
\langle \varepsilon({\bf x})\varepsilon({\bf y})\rangle -
\langle \varepsilon({\bf x})\rangle \langle \varepsilon({\bf y})\rangle
&=& \frac{g^4}{2}N^2(N^2-1) D(0)^2
\left[ D({\bf x}-{\bf y})^2 + E({\bf x}-{\bf y})^2 + F({\bf x}-{\bf y})^2 \right]
\nonumber \\
&+& \frac{7g^4}{16}N^2(N^2-1) 
\left[ D({\bf x}-{\bf y})^2 + E({\bf x}-{\bf y})^2 + F({\bf x}-{\bf y})^2 \right]^2 .
\end{eqnarray}
Since the functions $D, E, F$ always appear in the same combination, it makes sense to introduce the abbreviation
\begin{equation}
K({\bf x}-{\bf y}) = D({\bf x}-{\bf y})^2 + E({\bf x}-{\bf y})^2 + F({\bf x}-{\bf y})^2 ,
\end{equation}
In terms of which the average deposited energy density and its fluctuation can be expressed as:
\begin{eqnarray}
\varepsilon_0 &=& \langle \varepsilon \rangle = \frac{g^2}{2} N(N^2-1)K(0) ;
\\
\Delta\varepsilon({\bf x}-{\bf y})^2 &=& \langle \varepsilon({\bf x})\varepsilon({\bf y})\rangle -
\langle \varepsilon({\bf x})\rangle \langle \varepsilon({\bf y})\rangle
= \frac{g^4}{2}N^2(N^2-1) \left[ K(0)K({\bf x}-{\bf y}) + \frac{7}{8} K({\bf x}-{\bf y})^2 \right] .
\end{eqnarray}

\section{Evaluation of integrals}

Next we simplify the integrals $D({\bf x}-{\bf y})$, $E({\bf x}-{\bf y})$ and $F({\bf x}-{\bf y})$.
We abbreviate ${\bf z} = {\bf x}-{\bf y}$ and $z = |{\bf z}|$. We begin with $D({\bf z})$.
\begin{eqnarray}
D({\bf z})&=& \int_0^{\infty} \frac{p\, dp}{4\pi^2}\, G(p) 
~\int_0^{2\pi} d\phi\, \cos(pz_1\cos \phi )\, \cos(pz_2\sin \phi )  
\nonumber \\
&=& \int_0^{\infty} \frac{p\, dp}{4\pi^2}\, G(p) 
~\int_0^{\pi}d\phi\, \Big(\cos[p(z_1\cos \phi +z_2\sin \phi)]  
+ \cos[p(z_1\cos \phi -z_2\sin \phi)]\Big)  
\end{eqnarray}
We substitute $\phi \rightarrow \pi-\phi$ in the last term and introduce the notation $z_1=z\cos \psi$, $z_2=z\sin \psi$:  
\begin{eqnarray}
D({\bf z})&=& \int_0^{\infty} \frac{p\, dp}{4\pi^2}\, G(p) ~\int_0^{2\pi} d\phi~\, \cos[pz\cos(\phi-\psi)] 
\nonumber \\
&=& 
\int_0^{\infty} \frac{p\, dp}{2\pi}\, G(p) ~J_0(pz) 
= 
\int_0^{\infty} \frac{d^2p}{4\pi^2}\, G(p) ~e^{{\rm i}{\bf p}\cdot {\bf z}}
\equiv G(z) ,
\end{eqnarray}
where we have used Eq.~(3.715.18) from \cite{GR}. Similarly we obtain 
\begin{eqnarray}
E({\bf z})&=& \int_0^{\infty} \frac{p\, dp}{4\pi^2}\, G(p) ~\int_0^{2\pi} d\phi\, \cos(pz_1\cos \phi )\,
 \cos(pz_2\sin \phi )\, (\cos^2\phi -\sin^2\phi) 
\nonumber \\
&=& \int_0^{\infty} \frac{\, dp}{4\pi^2}\, G(p) ~\int_0^{\pi}d\phi\, 
\Big(\cos[p(z_1\cos \phi +z_2\sin \phi)] 
+ \cos[p(z_1\cos \phi -z_2\sin \phi)] \Big)\cos(2\phi)  .
\end{eqnarray}
Using the same substitutions we find:  
\begin{eqnarray}
E({\bf z})&=& \int_0^{\infty} \frac{p\, dp}{4\pi^2}\, G(p) ~\int_0^{2\pi} d\phi\, 
 \cos[pz\cos(\phi-\psi)] \cos(2\phi)
\nonumber \\
&=& -\cos(2\psi)\int_0^{\infty} \frac{p\, dp}{2\pi}\, G(p) J_2(pz) 
\end{eqnarray}
where we used Eqs.~(3.715.18) and (3.715.7) from \cite{GR}.
Finally, a similar calculation yields:
\begin{eqnarray}
F({\bf z})&=& \int_0^{\infty} \frac{p\, dp}{4\pi^2}\, G(p) ~\int_0^{2\pi} d\phi\, 
\sin(pz_1\cos \phi )\, \sin(pz_2\sin \phi )\cos\phi \sin\phi 
\nonumber \\
&=& \sin(2\psi)\int_0^{\infty} \frac{p\, dp}{2\pi}\, G(p) J_2(pz) .
\end{eqnarray}
\end{widetext}
We conclude that the function $K(z)$ only depends on the distance $z$ between the points ${\bf x}$ and ${\bf y}$.

We can express $D(z)$ and $E(z)^2 + F(z)^2$ in terms of the following integrals:
\begin{equation}
C_n(z) = \int_0^\infty \frac{p\, dp}{2\pi}\, G(p) J_n(pz) ,
\end{equation}
for $n=0,2$, namely:
\begin{equation}
D(z) = C_0(z) ; 
\end{equation}
\begin{equation}
E(z)^2 + F(z)^2 = C_2(z)^2 .
\end{equation}
We rewrite the integrals as follows:
\begin{eqnarray}
C_n(z) &=& \int \frac{d^2p}{(2\pi)^2} \int d^2x\, e^{-i{\bf p}\cdot{\bf x}} G({\bf x}) J_n(pz)
~~.
\nonumber \\
\end{eqnarray}
We have to evaluate integrals of the type
\begin{equation}
B_n(x,z) = \int_0^\infty \frac{p\, dp}{2\pi} J_0(px) J_n(pz) .
\end{equation}
This integral can be evaluated for $n=0$ using formula (6.633.2) in \cite{GR}:
\begin{equation}
\int_0^\infty p\, dp\, e^{-c^2p^2} J_0(px) J_0(pz) 
= \frac{e^{-(x^2+z^2)/4c^2}}{2c^2} I_0\Big(\frac{xz}{2c^2}\Big) .
\end{equation}
We are interested in the limit $c\to 0$, which means that we can apply the
limit of $I_0(z)$ for large arguments:
\begin{equation}
I_0(z) \to \frac{e^z}{\sqrt{2\pi z}} .
\end{equation}
This yields
\begin{eqnarray}
B_0(x,z) &=& \lim_{c\to 0} \frac{1}{4\pi c} \frac{1}{\sqrt{\pi xz}} \exp\Big(-\frac{(x-z)^2}{4c^2}\Big)
\nonumber \\
&=& \frac{1}{2\pi z} \delta(x-z) .
\end{eqnarray}

For $n=2$ we use the recursion relation for Bessel functions:
\begin{equation}
J_2(z) = \frac{2}{z} J_1(z) - J_0(z) ,
\end{equation}
and apply formula (6.512.3) from \cite{GR}:
\begin{equation}
\int_0^\infty dp\, J_0(px) J_1(pz) = \frac{1}{z} \theta(z-x) ,
\end{equation}
with the convention $\theta(0)=1/2$. This implies:
\begin{equation}
B_2(x,z) = \frac{1}{\pi z^2}\theta(z-x) - \frac{1}{2\pi z} \delta(x-z) .
\end{equation}

When we insert these results into the desired integrals, we find:
\begin{eqnarray}
C_0(z) &=& G(z) ;
\\
C_2(z) &=& \frac{2}{z^2} \int_0^z x\, dx\, G({\bf x}) - G(z) .
\end{eqnarray}

\section{Numerical results}

\begin{figure}[htb]   
\includegraphics[width=0.95\linewidth]{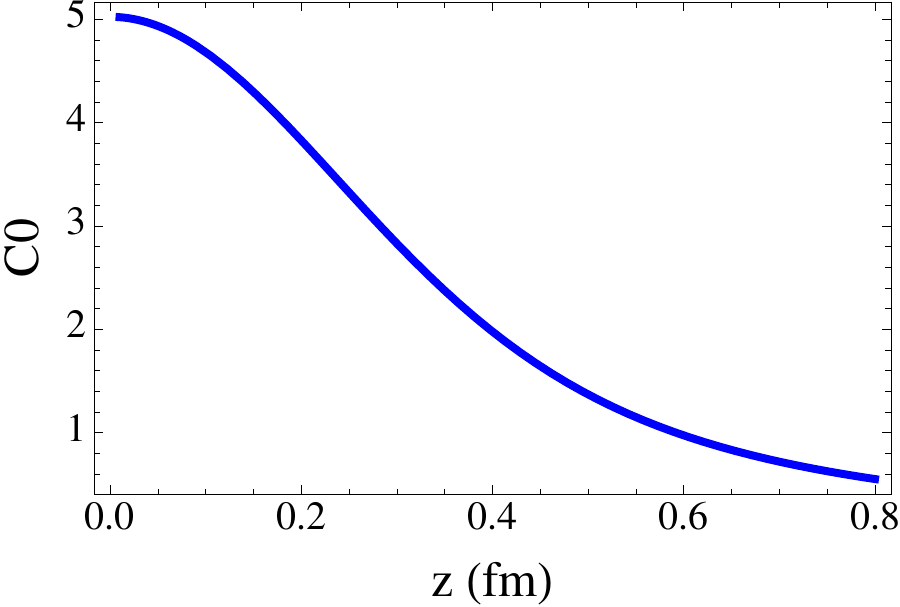}
\caption{The function $C_0(z)$ for the selected parameters.}
\label{fig:C0}
\end{figure}

\begin{figure}[htb]   
\includegraphics[width=0.95\linewidth]{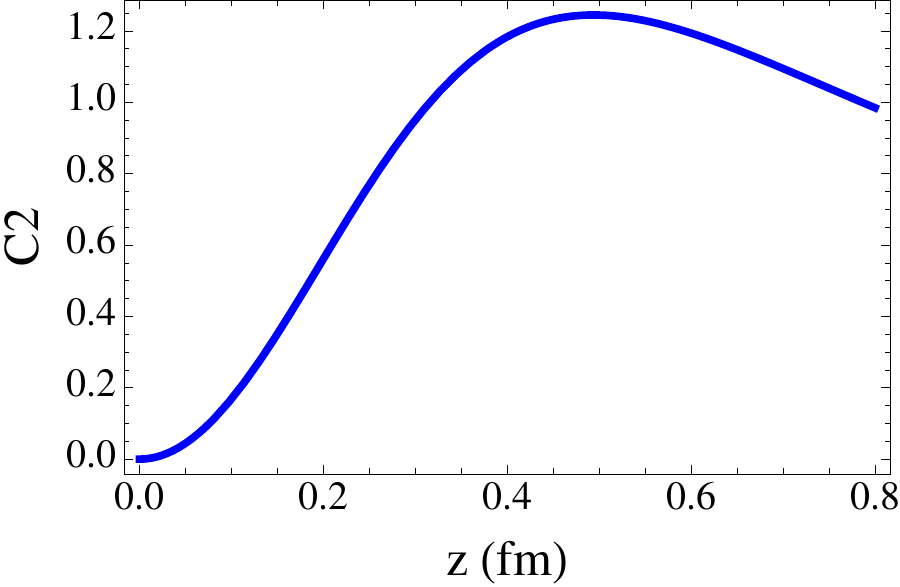}
\caption{The function $C_2(z)$ for the selected parameters.}
\label{fig:C2}
\end{figure}
We now evaluate the average energy density and its fluctuations for a choice of the parameters that is motivated by the initial conditions at which thermal QCD matter is formed in heavy ion collisions at RHIC and LHC:
\begin{eqnarray}
Q_s^2 &=& (g^2\mu)^2 = 2~{\rm GeV}^2 ;
\nonumber \\
g^2(\mu^2) &=& 3.785 ;
\nonumber \\
g^2(1/x^2) &=& \frac{16\pi^2}{9 \ln(1/(\Lambda^2 x^2))} .
\end{eqnarray}
Note that the result is independent of the value of $\Lambda$. The functions $C_0(z)$ and $C_2(z)$ are shown in Figs.~\ref{fig:C0} and \ref{fig:C2} for these parameter values.

\begin{figure}[htb]   
\includegraphics[width=0.95\linewidth]{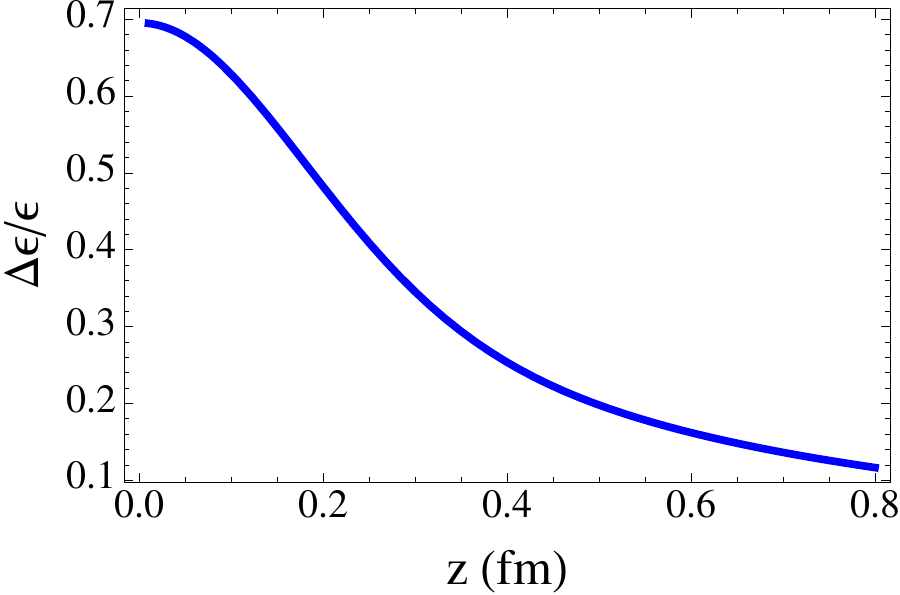}
\caption{The function $\Delta\varepsilon(z)/\varepsilon_0$ for the selected parameters.}
\label{fig:Deps}
\end{figure}

For these parameters, the initial value of the deposited energy density is $\varepsilon_0 \approx 240~{\rm GeV/fm}^3$. This very large energy density quickly decreases due to the longitudinal expansion and reaches much smaller values by the time of thermalization. What matters for us is not the absolute value of the initial energy density, but the relative size and spatial correlation of its fluctuations, $\Delta\varepsilon({\bf x}-{\bf y})/\varepsilon_0$. This function is shown in Fig.~\ref{fig:Deps} for the parameters listed above. 

As the figure shows, the fluctuations of the initial energy density are locally of similar magnitude as the energy density itself and fall over distances of the inverse saturation scale, here assumed as $Q_s^{-1} \approx 0.14~{\rm fm}$. This result is in accord with the intuitive picture of the field configuration immediately after the collision in the color glass condensate model, as a bundle of longitudinally stretching random color flux tubes with characteristic transverse width $1/Q_s$.

\section{Summary}

We have calculated the initial energy density fluctuations in high-energy heavy ion collisions within the Gaussian color glass condensate model. These turn out to be very large with a transverse profile determined by the saturation scale $Q_s$. A finite result is only obtained when the ``triumvirate'' running coupling is used, giving additional support to the correctness of Eq. (13). The fluctuation probabilities thus derived can serve as input for any calculation aiming at the investigation of early fluctuations, in particular for calculations which study the fate of such fluctuations during thermalization. For example, it is possible to investigate
the problem of event-by-event fluctuations in heavy ion collisions within the AdS/CFT paradigm using methods similar to those employed in \cite{string}.

{\em Acknowledgments:} This work was supported in part by the Office of Science of the U.S. Department of Energy (DE-FG02-05ER41367) and the BMBF (06 RY9191). We thank U.~Heinz, J.~Liu, Z.~Qiu, and C.~Shen for pointing out several typographical and numerical errors in an earlier version of this manuscript, and R.~Venugopalan and H.~Weigert for clarifying the  Gaussian approximation in the color glass condensate model.

\end{document}